\documentclass{article}

\usepackage{authblk} 
\usepackage{rotating} 
\usepackage{pdflscape}
\usepackage{paracol} 
\usepackage[numbers,sort&compress]{natbib} 
\usepackage{amssymb}
\usepackage{graphicx}
\usepackage{amsmath}
\usepackage{amsthm}
\usepackage{geometry}
\usepackage{enumitem}
\usepackage{algorithm}
\usepackage{algorithmic}
\usepackage{xcolor}
\usepackage{booktabs}
\usepackage{inputenc}
\usepackage{hyperref}
\usepackage[all,cmtip]{xy}

\theoremstyle{plain}

\theoremstyle{definition}

\begin{document}

\title{\textbf{Efficient Single-Stage Stellarator Optimization via a Modified Alternating Direction Method of Multipliers}}
\date{}
\author[1]{Junzhe Liu}
\author[1]{Haotian Chen}
\author[1]{Xishuo Wei}
\affil[1]{Future Energy Institute, Shanghai Jiao Tong University}

\maketitle

\begin{abstract}
    Designing stellarator magnetic fields with favorable confinement properties and practical coils is a challenging task. This design problem can be formulated as either a two-stage or a single-stage numerical optimization problem. The two-stage approach usually fails to satisfy coil engineering constraints while maintaining a sufficient mismatch between the coils and the target magnetic field. The single-stage optimization can suffer from convergence difficulties and high computational cost. This work proposes an efficient single-stage optimization algorithm based on a modified alternating direction method of multipliers (ADMM). By exploiting the variable-separable structure of the objectives, the single-stage problem can be decomposed into an equilibrium subproblem and a coil subproblem, coupled through constraints on the boundary normal field error and plasma-coil separation. This formulation allows existing equilibrium and coil optimization solvers to be incorporated into a common framework. Applications to quasi-isodynamic stellarator design demonstrate favorable trade off among physical constraints, coil complexity, and the computational cost compared with quasi-single-stage and single-stage methods.
\end{abstract}
\section{Introduction}

Stellarators offer a promising route to steady-state fusion operation because they can confine plasmas without relying on a large toroidal plasma current, reducing their susceptibility to current-driven disruptions. Their three-dimensional magnetic geometry provides considerable freedom to tailor plasma confinement and stability, but also poses a demanding design problem: favorable magnetic fields must be realized by external coils that satisfy practical engineering constraints. These requirements are closely coupled, since a configuration with excellent plasma properties may require complex coils, while simplifying the coils can compromise the magnetic field they produce. Stellarator design is therefore naturally formulated as a numerical optimization problem that seeks to reconcile plasma performance, coil feasibility, and the accuracy with which the coils reproduce the target field.

The design of a stellarator is conventionally cast as a two-stage optimization problem~\cite{liu_optimizing_2026, velasco_piecewise_2024, landreman_omnigenity_2012, liu_optimization_2026, helander_theory_2014, dudt_magnetic_2024, goodman_quasi-isodynamic_2024, helander_bootstrap_2009, padidar_diffusion_2025, gil_stellarator_2026, kruger_coil_2025, yamaguchi_optimization_2021, wu_planar_2025, nie_focus-hts_2025, hammond_framework_2025, zhu_designing_2018, gil_efficient_2026}. In the first stage, an MHD equilibrium is obtained by solving the force balance equations together with a set of physics targets, such as low transport or favorable stability properties~\cite{liu_optimizing_2026, velasco_piecewise_2024, landreman_omnigenity_2012, liu_optimization_2026, helander_theory_2014, dudt_magnetic_2024, goodman_quasi-isodynamic_2024, helander_bootstrap_2009, padidar_diffusion_2025}. In the second stage, the boundary of that equilibrium is taken as a fixed target, and an inverse magnetostatics problem is solved to find coils whose field is tangential to it~\cite{gil_stellarator_2026, kruger_coil_2025, yamaguchi_optimization_2021, wu_planar_2025, nie_focus-hts_2025, hammond_framework_2025, zhu_designing_2018, gil_efficient_2026}. This two-stage structure is attractive because each subproblem can be solved with mature numerical tools, and it is for this reason that the majority of stellarator design studies to date have followed it. The decoupling, however, comes at a cost: the coils are optimized against a boundary that was determined without any knowledge of them, so the final plasma-coil pair is not required to be mutually consistent.

An alternative is to treat the plasma boundary and the coils as variables of a single-stage optimization problem and solve for them jointly~\cite{henneberg_combined_2021, giuliani_single-stage_2022, jorge_single-stage_2023, jorge_simplified_2024}. In this single-stage formulation, the magnetic surface and the coil geometry are allowed to adjust to each other during the optimization, so that the resulting configuration can be internally consistent by construction rather than by post hoc reconciliation. This is a genuine advantage that the two-stage approach cannot offer, and it motivates the growing interest in single-stage stellarator optimization despite the fact that relatively few such studies exist so far. \cite{giuliani_single-stage_2022} proposed a single-stage coil design approach that directly optimizes coils for a vacuum quasi-symmetric field and demonstrates that quasi-Newton methods are crucial for convergence and that the flat landscape allows simple, manufacturable coils. \cite{jorge_single-stage_2023} presented a method that jointly optimizes the plasma boundary and coils under fixed-boundary equilibrium through a quadratic flux penalty. \cite{jorge_simplified_2024} illustrated that combined plasma-coil optimization streamlines stellarator design, yielding new configurations with very few coils.

One of the difficulties within the single-stage approach lies in its computational cost. Coupling the equilibrium and the coils introduces a large number of variables and substantially enlarges the search space~\cite{giuliani_single-stage_2022, jorge_single-stage_2023, jorge_simplified_2024}. The resulting high-dimensional and complicated landscape makes it difficult to locate a good solution, and the computational effort required to do so grows accordingly, i.e. the evaluation of the objective and its derivatives, together with the associated matrix operations, becomes considerably more expensive. As a result, even a single optimization loop requires a large computational effort, which makes the single-stage approach slow in practice. In addition, the objectives and constraints in stellarator design are generally nonlinear, nonconvex and noncontinuous, which further degrades the convergence and stability of the single-stage optimization~\cite{conlin_stellarator_2024}. To make single-stage optimization efficient, stable and even robust, without weakening its physical fidelity, is therefore an important step toward its practical use.

In this work, we propose an efficient single-stage algorithmic framework for stellarator optimization. Motivated by distributed optimization algorithms~\cite{bolte_proximal_nodate, boyd_distributed_2010, eckstein_douglasrachford_1992, bot_proximal_2020, li_global_2015, wang_global_2018, bian_stochastic_2021, cui_convergence_2015}, plug-and-play manner~\cite{chan_plug-and-play_2016, ryu_plug-and-play_2019}, acceleration techniques~\cite{el_bourkhissi_convergence_2026, cui_convergence_2015, bian_stochastic_2021} and convergence analysis~\cite{hien_multiblock_2023, el_bourkhissi_convergence_2026, bian_stochastic_2021, bot_proximal_2020} from the mathematical optimization literature, we exploit the mathematical structure of the objective in single-stage stellarator optimization. Accordingly, the proposed framework effectively addresses both the high per-iteration computational cost and the instability of the conventional single-stage approach. Applying the proposed method reduces the computational time and improves the stability while preserving the physical performance, enabling us to explore a range of stellarator designs.
\section{Optimization Model with Separated Objective Form}

\subsection{Preliminaries}
In stellarator optimization, the quantities that are minimized or constrained fall into two broad categories: the plasma equilibrium and the external coils. Both the plasma boundary and the coil geometry are usually represented by truncated Fourier series, which provide a smooth and compact parameterization well suited to optimization. 

The plasma boundary, denoted by $\partial \Omega$, is described by a Fourier--RZ toroidal surface,
$$
    R(\theta,\zeta) = \sum_{m,n} R_{mn} \cos(m\theta - n\zeta), \qquad
    Z(\theta,\zeta) = \sum_{m,n} Z_{mn} \sin(m\theta - n\zeta),
$$
where $\theta$ and $\zeta$ are the poloidal and toroidal angles, $R_{mn}$ and $Z_{mn}$ are the Fourier coefficients, and $n$ is restricted to integer multiples of the number of field periods so that the surface respects the toroidal symmetry of the device. The full set of equilibrium design variables, collectively denoted by $x_{\mathrm{eq}}$, comprises the boundary Fourier coefficients $\{R_{mn}, Z_{mn}\}$, the pressure and current profile parameters, the rotational transform profile, and the total toroidal flux:
$$
    x_{\mathrm{eq}} := \bigl\{ R_{mn},\, Z_{mn},\, p_l,\, c_l,\, \iota_l,\, \Psi \bigr\}.
$$

Similarly, each coil is represented as a closed three-dimensional curve in Cartesian coordinates~\cite{zhu_new_2018},
$$
    \mathbf{r}(t) = \sum_{k=0}^{K} \Bigl[ \mathbf{a}_k \cos(k t) + \mathbf{b}_k \sin(k t) \Bigr],
    \qquad t \in [0, 2\pi),
$$
where $\mathbf{a}_k, \mathbf{b}_k \in \mathbb{R}^3$ are the Fourier coefficients of the curve and $K$ is the truncation order. The coil current, rotation matrix, and rigid shift are treated as additional parameters. The complete set of coil parameters, collectively denoted by \(x_{\mathrm{coil}}\), is therefore
$$
    x_{\mathrm{coil}} := \bigl\{ \mathbf{a}_k,\, \mathbf{b}_k,\, I,\, \mathbf{R},\, \mathbf{s} \bigr\},
$$
where \(I\) is the coil current, \(\mathbf{R}\) is the rotation matrix, and \(\mathbf{s}\) is the rigid shift.

\subsection{Classification of objective functions}
The objectives used in magnetic equilibrium and coil design are numerous, and a comprehensive review is beyond the scope of this paper. Broadly, equilibrium optimization includes geometric targets, omnigenity, quasi-symmetry metrics, MHD stability proxies and fast-particle confinement measures. Coil optimization usually involves normal field error, coil-coil distance, plasma-coil distance, curvature and length, subject to numerous engineering constraints in practice. According to the variables on which each objective function depends, we divide them into the following three categories and show them in Table~\ref{objective table}.

\emph{Objectives depending on equilibrium only}. In the model, we focus on a representative subset consisting of omnigenity error $f_{\text{QI}}$, effective ripple $\epsilon_{\text{eff}}$ and other terms listed in Table~\ref{objective table}. All of these objectives that depend only on equilibrium are ultimately collected into an equilibrium objective function $f_{eq}(x_{eq})$ with predefined weights. The dominant requirement is the MHD force balance condition; the conditions that fix the boundary shape, the pressure and current profiles, and the total toroidal flux are listed among the constraints. These constraints ensure that the optimized equilibrium remains a genuine solution of the MHD equilibrium problem rather than an arbitrary field configuration.

\emph{Objectives depending on coils only}. For the engineering constraints, we consider coil distance, curvature, torsion and other objectives listed in Table~\ref{objective table}. Each of these requirements can be expressed as an indicator function, which is zero when the corresponding constraint is satisfied and infinite otherwise. These terms depending on coils only are collected into a coil objective $f_{coil}(x_{coil})$ with predefined weights. In this way, nearly all the engineering constraints are handled within a single optimization formulation. This structure is favorable for the optimizer, since each constraint is encoded as an indicator function.

\emph{Objectives depending on both equilibrium and coils}. There are usually two objectives that contain both $x_{eq}$ and $x_{coil}$. One is normal field error, which treats the plasma boundary and the coils as variables so that the two are required to be mutually consistent. This consistency is expressed through the condition
\begin{equation}\label{constraint}
    c(x_{eq}, x_{coil}):= \int_{\partial \Omega} \frac{1}{2}\left(\mathbf{B}_{coil} \cdot \hat{\mathbf{n}}\right)^2 ds = 0,
\end{equation}
where \(\mathbf{B}_{coil}\) denotes the magnetic field produced by the external coils and \(\hat{\mathbf{n}}\) is the outward unit normal on the boundary \(\partial\Omega\). This condition states that the coil field has no normal component on \(\partial\Omega\), i.e., the boundary is a flux surface of the coil field. Enforcing it within the same optimization that determines the boundary is what distinguishes the single-stage approach from the two-stage one, in which the boundary is fixed in advance and the condition is only approximately satisfied a posteriori. The other is minimum plasma-coil distance $f_{cp}$, namely 
\begin{equation}\label{fcp}
    f_{cp}(x_{\mathrm{eq}}, x_{\mathrm{coil}})=
    \begin{cases}
        0, & \text{if the minimum plasma--coil distance exceeds a prescribed threshold}\\
        +\infty, & \text{otherwise}
    \end{cases}
\end{equation}

\begin{table}[H]
\centering
\caption{Objectives and constraints in the proposed model.}
\label{objective table}
\begin{tabular}{@{}p{0.30\textwidth}p{0.66\textwidth}@{}}
\toprule
\textbf{Objective / Constraint} & \textbf{Definition} \\
\midrule
\multicolumn{2}{@{}l}{\emph{Terms depend only on equilibrium}} \\
\midrule
Omnigenity error $f_{\mathrm{QI}}$
  & Deviation of the field strength on the flux surfaces from an omnigenous
    field, $\sum_\rho \int (B - B_{\mathrm{omni}})^2\,\mathrm{d}S$; drives
    trapped-particle confinement. \\
Effective ripple $\varepsilon_{\mathrm{eff}}$
  & Neoclassical transport proxy, $\varepsilon_{\mathrm{eff}}^{3/2}$ evaluated
    on several flux surfaces. \\
Mercier criterion $D_{\mathrm{Merc}}$~\cite{glasser_generalized_2023}
  & Local ideal-MHD stability; $D_{\mathrm{Merc}} > 0$ is required. \\
Magnetic well $W$
  & Global stability; $W > 0$ is favorable. \\
Shear $s$
  & Magnetic shear of the field. \\
Ballooning growthrate $\lambda_{\mathrm{ball}}$
  & Pressure-driven ideal instability; $\lambda_{\mathrm{ball}} < \lambda_0$ is
    required. \\
Mirror ratio $\Delta$
  & $\Delta = (B_{\max}-B_{\min})/(B_{\max}+B_{\min})$ on the last closed flux
    surface. \\
Aspect ratio $A$
  & $A = R_0 / a$. \\
Elongation $E$
  & $E = a_{\mathrm{major}} / a_{\mathrm{minor}}$. \\
Rotational transform $\iota$
  & $\iota$ on axis and at the edge, kept away from low-order rationals. \\
Force balance (constraint)
  & $\nabla p = \mathbf{J} \times \mathbf{B}$, enforced as a hard constraint. \\
\midrule
\multicolumn{2}{@{}l}{\emph{Terms depend only on coils}} \\
\midrule
Coil length $L_i$ \ [m]
  & $L_i = \int |\mathbf{r}_i'(t)|\,\mathrm{d}t$; the amount of conductor. \\
Max curvature $\kappa_{\max}$ \ [m$^{-1}$]
  & $\kappa_{\max,i} = \max_t \kappa_i(t)$; inverse of the tightest bending
    radius. \\
Coil-coil distance $d_{\mathrm{cc}}$ \ [m]
  & $\min_{i\neq j,\,t,t'} |\mathbf{r}_i(t) - \mathbf{r}_j(t')|$; room for the
    winding packs and their support. \\
Weighted torsion $\bar{\tau}$ \ [m$^{-1}$]
  & $\bar{\tau}_i = \bigl( \int \kappa_i^2 \tau_i^2\,\mathrm{d}\ell /
    \int \kappa_i^2\,\mathrm{d}\ell \bigr)^{1/2}$; curvature-weighted RMS
    torsion of a coil. \\
\midrule
\multicolumn{2}{@{}l}{\emph{Terms depend on both equilibrium and coils}} \\
\midrule
Normal-field residual $B_n$
  & $B_n = \mathbf{B}\cdot\hat{\mathbf{n}}$ on the boundary $\Gamma$; the
    consensus condition $B_n \to 0$, monitored by $r=\langle|B_n|/B\rangle$. \\
Plasma--coil distance $d_{\mathrm{cs}}$ [m]
  & $\min_{i,\,x\in\Gamma} |\mathbf{r}_i - x|$; clearance for the vessel,
    blanket, and shielding. \\
\bottomrule
\end{tabular}
\end{table}

\subsection{Final mathematical model}
Before presenting the final mathematical model, we highlight two key differences from conventional formulations. First, instead of including the normal-field error $\int (\mathbf{B}\cdot\mathbf{n})^2$ in the objective with a fixed weight, which typically drives \(B_n\) to a small value but can render the engineering constraints unreliable, we treat it as a constraint. This allows its effective weight to adjust automatically and avoid the trade-off between \(B_n\) and the plasma/coil objectives. Second, the objective is shaped in a variable-separable form, which makes it amenable to a distributed optimization algorithm and thereby reduces the computational cost. The final mathematical model is therefore formulated as
\begin{equation}\label{model}
    \begin{aligned}
        \min_{x_{eq}, x_{coil}} & ~ f_{eq}(x_{eq})+f_{coil}(x_{coil})\\
        \text{s.t.} & ~ 
        \begin{cases}
            c(x_{eq}, x_{coil})=0\\
            f_{cp}(x_{eq}, x_{coil})=0
        \end{cases}
    \end{aligned}
\end{equation}
where $f_{eq}(x_{eq})$ contains the magnetic field optimization objective functions with variable $x_{eq}$ only and $f_{coil}(x_{coil})$ contains $x_{coil}$ only. Nonlinear constraints $f_{cp}(x_{eq}, x_{coil})$ and $c(x_{eq}, x_{coil})$ are defined by \eqref{constraint} and \eqref{fcp}.

\section{Algorithm Design}

\subsection{Exploiting the mathematical structure of the problem}
The single-stage formulation is computationally demanding, as discussed above. A closer inspection of the optimization problem, however, reveals that it possesses a separable structure in its variables: the equilibrium and the coil parameters enter the objective in a way that allows the problem to be decomposed into smaller subproblems. This observation motivates us to solve the problem in a distributed manner rather than as a single monolithic optimization. Accordingly, we adopt the alternating direction method of multipliers (ADMM)~\cite{boyd_distributed_2010}, a distributed optimization algorithm that exploits exactly this separability, thereby reducing the cost of each iteration and improving the overall efficiency of the single-stage approach. The resulting speedup will be demonstrated in the experimental section.

In addition, the term $f_{coil}(x_{coil})$ in the objective is an indicator function, which can be interpreted as a projection onto the corresponding constraint set. This observation connects our formulation to a body of work on mathematical convergence analysis, where proximal or projection-based approaches for such terms are known to yield favorable convergence properties~\cite{bian_stochastic_2021, wang_global_2018, bot_proximal_2020}. This connection motivates us to exploit this mathematical structure, with favorable convergence properties. We also demonstrate this in the experimental section.

\subsection{Modified ADMM for single-stage stellarator optimization}
With the special mathematical structure discussed above, we find it useful to apply modified ADMM to tackle this optimization problem. Consequently, the augmented Lagrangian function is defined by
$$\begin{aligned}
    \hat{\mathcal{L}}_r(x_{eq}, x_{coil}, \lambda, \mu) := & f_{eq}(x_{eq})+f_{coil}(x_{coil})+ \langle \lambda, c(x_{eq}, x_{coil}) \rangle + \langle \mu, f_{cp}(x_{eq}, x_{coil}) \rangle \\
    & + \frac{r}{2}(\|c(x_{eq}, x_{coil})\|^2_2 + \|f_{cp}(x_{eq}, x_{coil})\|_2^2)
\end{aligned}$$
where $\lambda$ and $\mu$ are Lagrangian multipliers. Since $f_{cp}(x_{eq}, x_{coil})$ is an indicator function, the coefficient $\mu$ and $\frac{r}{2}$ can be omitted. Therefore, we define the simplified augmented Lagrangian function by
\begin{equation}\label{Lr}
    \mathcal{L}_r(x_{eq}, x_{coil}, \lambda) := f_{eq}(x_{eq})+f_{coil}(x_{coil})+ f_{cp}(x_{eq}, x_{coil}) + \langle \lambda, c(x_{eq}, x_{coil}) \rangle + \frac{r}{2}\|c(x_{eq}, x_{coil})\|^2_2 
\end{equation}
Thus, the k-th iteration scheme of the proposed ADMM can be expressed as
\begin{subequations} \label{iteration}
    \begin{align}
    & x_{eq}^{k+1}=\underset{x_{eq}}{\text{argmin}} ~\mathcal{L}_r(x_{eq}, x_{coil}^k, \lambda^k) \label{iteration_a}\\
    & x_{coil}^{k+1}=\underset{x_{coil}}{\text{argmin}} ~\mathcal{L}_r(x_{eq}^{k+1}, x_{coil}, \lambda^k)  \label{iteration_b}\\
    & \lambda^{k+1}=\lambda^k + r\cdot c(x_{eq}^{k+1}, x_{coil}^{k+1}) \label{iteration_c}
    \end{align}
\end{subequations}
According to \eqref{Lr}, \eqref{iteration_a} can be equivalently expressed as
\begin{equation}\label{iteration_eq}
    x_{eq}^{k+1} =\underset{x_{eq}}{\text{argmin}}~ f_{eq}(x_{eq})+f_{cp}(x_{eq}, x_{coil}^k)+\langle \lambda^k, c(x_{eq}, x_{coil}^k) \rangle + \frac{r}{2}\|c(x_{eq}, x_{coil}^k)\|^2_2
\end{equation}
Substituting \eqref{constraint} into \eqref{iteration_eq}, the third and forth terms yields
$$
    \lambda^k \cdot \int_{\partial \Omega} \frac{1}{2}\left(\mathbf{B}_{coil}(x_{coil}^k) \cdot \hat{\mathbf{n}}\right)^2 ds + \frac{r}{2}\|\int_{\partial \Omega} \frac{1}{2}\left(\mathbf{B}_{coil}(x_{coil}^k) \cdot \hat{\mathbf{n}}\right)^2 ds\|^2_2
$$
which is computationally expensive. Therefore, we replace it with a coil-fixed $B_n$ term. Then iteration \eqref{iteration_a} yields
$$
    x_{eq}^{k+1} = \underset{x_{eq}}{\text{argmin}} ~ f_{eq}(x_{eq})+f_{cp}(x_{eq}, x_{coil}^k) + (\lambda^k+\sqrt{\frac{r}{2}}) \cdot \text{coil-fixed $B_n$ term}
$$
Since $f_{cp}$ is an indicator function, it motivates the use of proximal-based methods~\cite{bot_proximal_2020, eckstein_douglasrachford_1992}, in which the smooth part of the objective is handled by a gradient step or trust region step and the non-smooth terms are handled exactly through their proximal operators, i.e. projections onto the corresponding constraint sets. Thus we let
$$
    F_1(x_{eq}, x_{coil}^k):= f_{eq}(x_{eq})+ (\lambda^k+\sqrt{\frac{r}{2}}) \cdot \text{coil-fixed $B_n$ term}
$$
for notational convenience in what follows. Similarly, \eqref{iteration_b} is equivalent to
\begin{equation}\label{iteration_coil}
    x_{coil}^{k+1} =\underset{x_{coil}}{\text{argmin}} ~ f_{coil}(x_{coil})+f_{cp}(x_{eq}^{k+1}, x_{coil}) + \langle \lambda^k, c(x_{eq}^{k+1}, x_{coil}) \rangle + \frac{r}{2}\|c(x_{eq}^{k+1}, x_{coil})\|^2_2
\end{equation}
Substituting \eqref{constraint} into \eqref{iteration_coil}, the third and forth terms yields
$$
    \lambda^k \cdot \int_{\partial \Omega} \frac{1}{2}\left(\mathbf{B}_{coil}(x_{coil}) \cdot \hat{\mathbf{n}}\right)^2 ds + \frac{r}{2}\|\int_{\partial \Omega} \frac{1}{2}\left(\mathbf{B}_{coil}(x_{coil}) \cdot \hat{\mathbf{n}}\right)^2 ds\|^2_2
$$
which can be replaced by a eq-fixed $B_n$ term. Then \eqref{iteration_b} can be rewritten as
$$
    x_{coil}^{k+1}=\underset{x_{coil}}{\text{argmin}} ~ f_{coil}(x_{coil})+f_{cp}(x_{eq}^{k+1}, x_{coil}) + (\lambda^k+\sqrt{\frac{r}{2}}) \cdot \text{eq-fixed $B_n$ term}
$$
Since both $f_{coil}$ and $f_{cp}$ are indicator functions, we let
$$
    F_2(x_{eq}^{k+1}, x_{coil}):= (\lambda^k+\sqrt{\frac{r}{2}}) \cdot \text{eq-fixed $B_n$ term}
$$
for notational convenience. Finally, the algorithm can be expressed as
\begin{algorithm}[H]
    \caption{Modified ADMM for single-stage stellarator optimization (MASO)}
    \label{ADMM}
    \begin{algorithmic}
        \STATE Initialize $x_{eq}^0$, $x_{coil}^0$ and $\lambda^0$; choose $r>0$, the maximum number of iterations $K$ and step-size $\gamma$.
        \FOR{$k=0,\ldots,K-1$}
            \STATE Compute \eqref{iteration_a} via $$x_{eq}^{k+1} = \operatorname{prox}_{\gamma, f_{cp}}\left( x_{eq}^{k} - \gamma \mathcal{D}_{F_1}(x_{eq}^{k}) \right)$$
            \STATE Compute \eqref{iteration_b} via $$x_{coil}^{k+1} = \operatorname{prox}_{\gamma, (f_{coil}+f_{cp})}\left( x_{coil}^{k} - \gamma \mathcal{D}_{F_2}(x_{coil}^{k}) \right)$$
            \STATE Update multipliers $$\lambda^{k+1}=\lambda^k + r\cdot c(x_{eq}^{k+1}, x_{coil}^{k+1})$$
        \ENDFOR
        \STATE \textbf{Output:} $x_{eq}^K$ and $x_{coil}^K$.
    \end{algorithmic}
\end{algorithm}
The proposed MASO is not only a single monolithic algorithm but also a general plasma-coil co-optimization framework. As a framework, it can cooperate with existing stellarator optimization methods, namely VMEC~\cite{10.1063/1.864116}, SIMSOPT~\cite{landreman_simsopt_2021}, DESC~\cite{panici_desc_2023, conlin_desc_2023, dudt_desc_2023}, FOCUS~\cite{zhu_new_2018, zhu_designing_2018, nie_focus-hts_2025} and other approaches, in a plug-and-play manner~\cite{chan_plug-and-play_2016, ryu_plug-and-play_2019}. Thus, the most suitable solver can be selected for each subproblem in a given scenario since the two blocks communicate through the constraints. Each block can keep its own objectives and constraints in their natural variables, and the coupling is handled entirely by the multiplier. This explicit decoupling gives MASO a high degree of freedom and universality: it can accommodate a wide variety of physics and engineering objectives, new solvers and constraints, without reformulating the whole problem. We will demonstrate the efficiency of MASO and the plug-and-play capability in the following discussion.
\section{Applications for QI-Stellarator Design}
To assess the effectiveness of the proposed approach MASO, we compare it with three baseline optimization strategies, namely quasi-single-stage (QSS), single-stage with least squares (SS-lsq), single-stage with augmented Lagrangian method (SS-aug), while keeping the same physical objectives and initial conditions so that the comparison isolates the effect of the algorithm alone. 

The quasi-single-stage method models the coils as a surface current on a winding surface and optimizes the current potential jointly with the equilibrium. The single-stage method instead introduces a finite set of Fourier coils as explicit optimization variables and solves the entire problem in one pass. These method families span the range from fully decoupled to fully coupled equilibrium-coil optimization, against which the alternating formulation of this work is evaluated.

\begin{figure}[H]
    \centering
    \includegraphics[width=0.75\textwidth]{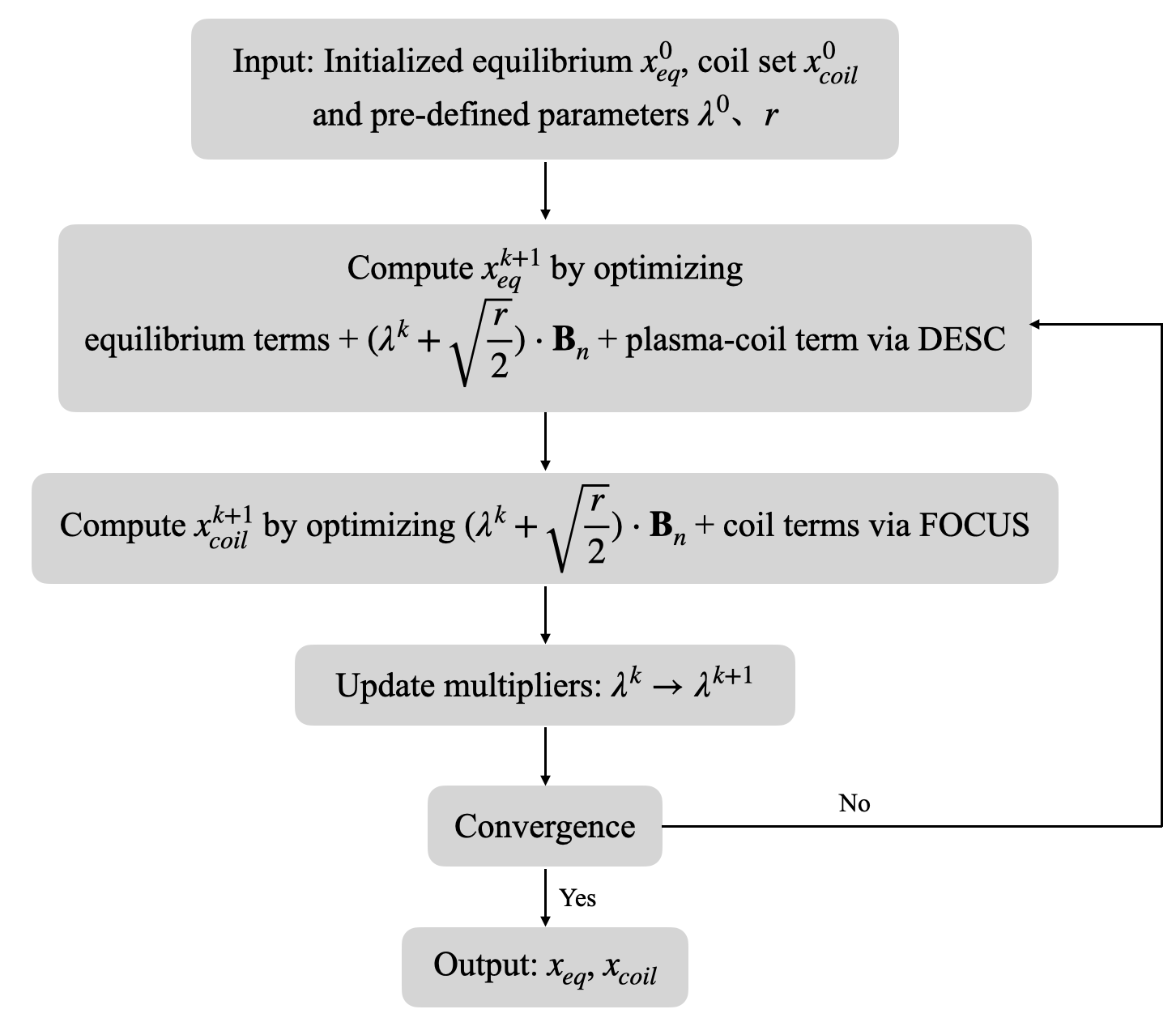}
    \caption{Illustration of plug-and-play MASO process for QI-stellarator}
    \label{algorithm for honghu}
\end{figure}

Owing to the plug-and-play nature of the framework, we adopt DESC to optimize the plasma boundary and FOCUS to optimize the coils. The details of the optimization process is shown in Figure~\ref{algorithm for honghu}. 

For the experimental settings, a four-field-period quasi-isodynamic (QI) stellarator with 32 modular coils and a major radius $R_0 = 2.0$ m is adopted. Exploiting the stellarator symmetry and the four-fold field periodicity, the coil set is parameterized by four independent coils, from which the remaining coils are generated by symmetry replication. We test these four approaches on a single NVIDIA GPU. The equilibrium objectives consist of omnigenity error, mercier stability, magnetic well, ballooning stability, effective ripple, mirror ratio, aspect ratio, elongation and rotational transform together with its shear, with force balance enforced as a constraint. The coil-side objectives consist of boundary normal field error, coil-coil and plasma-coil minimum distance, coil curvature, torsion and coil length. 
\begin{figure}[htbp]
    \centering
    \includegraphics[width=0.95\textwidth]{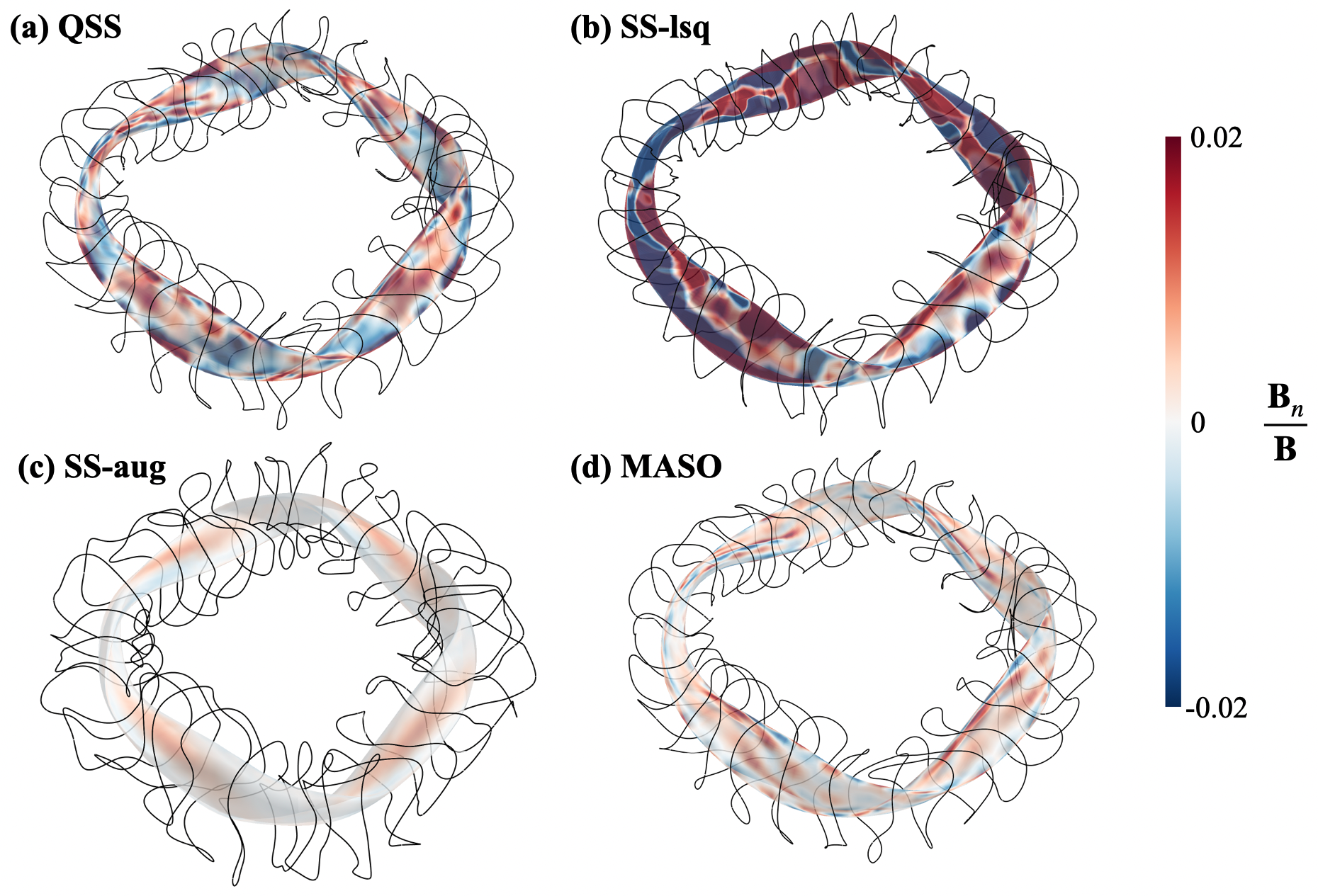}
    \caption{Plasma boundaries and optimized coil sets obtained by the four methods: QSS, SS-lsq, SS-aug and MASO.}
    \label{4-compare}
\end{figure}

\begin{table}[H]
\centering
\caption{Comparison of the objective values of QSS, SS-lsq, SS-aug and MASO.}
\label{results table}
\begin{tabular}{@{}lcccc@{}}
\toprule
 & \textbf{QSS} & \textbf{SS-lsq} & \textbf{SS-aug} & \textbf{MASO} \\
\midrule
\multicolumn{5}{@{}l}{\emph{Plasma performance}} \\
\midrule
Effective ripple $\varepsilon_{\mathrm{eff}}^{3/2}$  & $5.89\times10^{-4}$ & $7.73\times10^{-4}$ & $9.08\times 10^{-4}$ & $5.19\times10^{-4}$ \\
Mercier criterion $D_{\mathrm{Merc}}$ & >0 & >0 & >0 & >0 \\
Ballooning $\lambda_{\mathrm{ball}}$ & 0.48 & $2.01\times 10^{-3}$ & $-0.12\times 10^{-3}$ & 0.27\\
Mirror ratio $\Delta$ & 0.18 & 0.18 & 0.23 & 0.10 \\
Aspect ratio $A$ & 10.94 & 10.95 & 9.83 & 11.86 \\
Elongation $E$ & 6.16 & 6.38 & 5.74 & 6.12 \\
Rotational transform $\iota$ & 0.81 / 0.95 & 0.93/0.95 & 0.91/0.98 & 0.87 / 0.99 \\
\midrule
\multicolumn{5}{@{}l}{\emph{Coil feasibility}} \\
\midrule
$\langle |B_n|/B \rangle$ (mean) & $4.68\times10^{-3}$ & $1.30\times 10^{-2}$ & $8.17\times 10^{-4}$ & $3.62\times10^{-3}$ \\
Coil-coil distance $d_{\mathrm{cc}}$ \ [m] & 0.17 & 0.16 & 0.13 & 0.17 \\
Plasma-coil distance $d_{\mathrm{cs}}$ \ [m] & 0.24 & 0.22 & 0.18 & 0.26 \\
Coil length $L$ \ [m] & 4.17 & 3.50 & 5.71 & 4.06 \\
Max curvature $\kappa_{\max}$ \ [m$^{-1}$] & 4.39 & 2.73 & 7.18 & 4.38 \\
\midrule
\multicolumn{5}{@{}l}{\emph{Computational cost}} \\
\midrule
GPU time  & - & 7-10 h & 190-290h & 4-7h  \\
\bottomrule
\end{tabular}
\end{table}

Figure~\ref{4-compare} and Table~\ref{results table} illustrate the cases generated by four methods with the same initial settings. We first note that, after an extensive search over weight combinations, two of the four methods are not competitive. SS-lsq leaves an unacceptably large residual normal field ($\langle |B_n|/B\rangle = 1.30\times10^{-2}$), more than three times larger than the other methods. SS-aug achieves the smallest normal field error ($8.17\times10^{-4}$) but produces coils that violate the engineering limits: its coil-coil distance drops to $0.13$ m, its plasma-coil distance to $0.18$ m and its maximum curvature rises to $7.18$ m$^{-1}$. Moreover, SS-aug requires an extremely expensive GPU running time. We therefore exclude these two methods from the subsequent comparison and focus on QSS and MASO.

QSS and MASO reach comparable plasma performance, while MASO consistently improves the coil feasibility. The effective ripple is reduced from $5.89\times10^{-4}$ to $5.19\times10^{-4}$ and the mirror ratio is brought from $0.18$ down to the design target of $0.10$. On the coil side, MASO lowers the mean normal field error from $4.68\times10^{-3}$ to $3.62\times10^{-3}$, enlarges the plasma-coil clearance from $0.24$ to $0.26$ m and shortens the total coil length from $4.17$ to $4.06$ m, while keeping a comparable maximum curvature ($4.38$ vs $4.39$ m$^{-1}$) and the same coil-coil distance ($0.17$ m). The remaining quantities, Mercier stability, ballooning stability, aspect ratio, elongation and rotational transform, are essentially equivalent between the
two methods. These results demonstrate that MASO yields a more manufacturable coil set without sacrificing plasma performance relative to QSS.

\begin{figure}[H]
    \centering
    \includegraphics[width=0.95\textwidth]{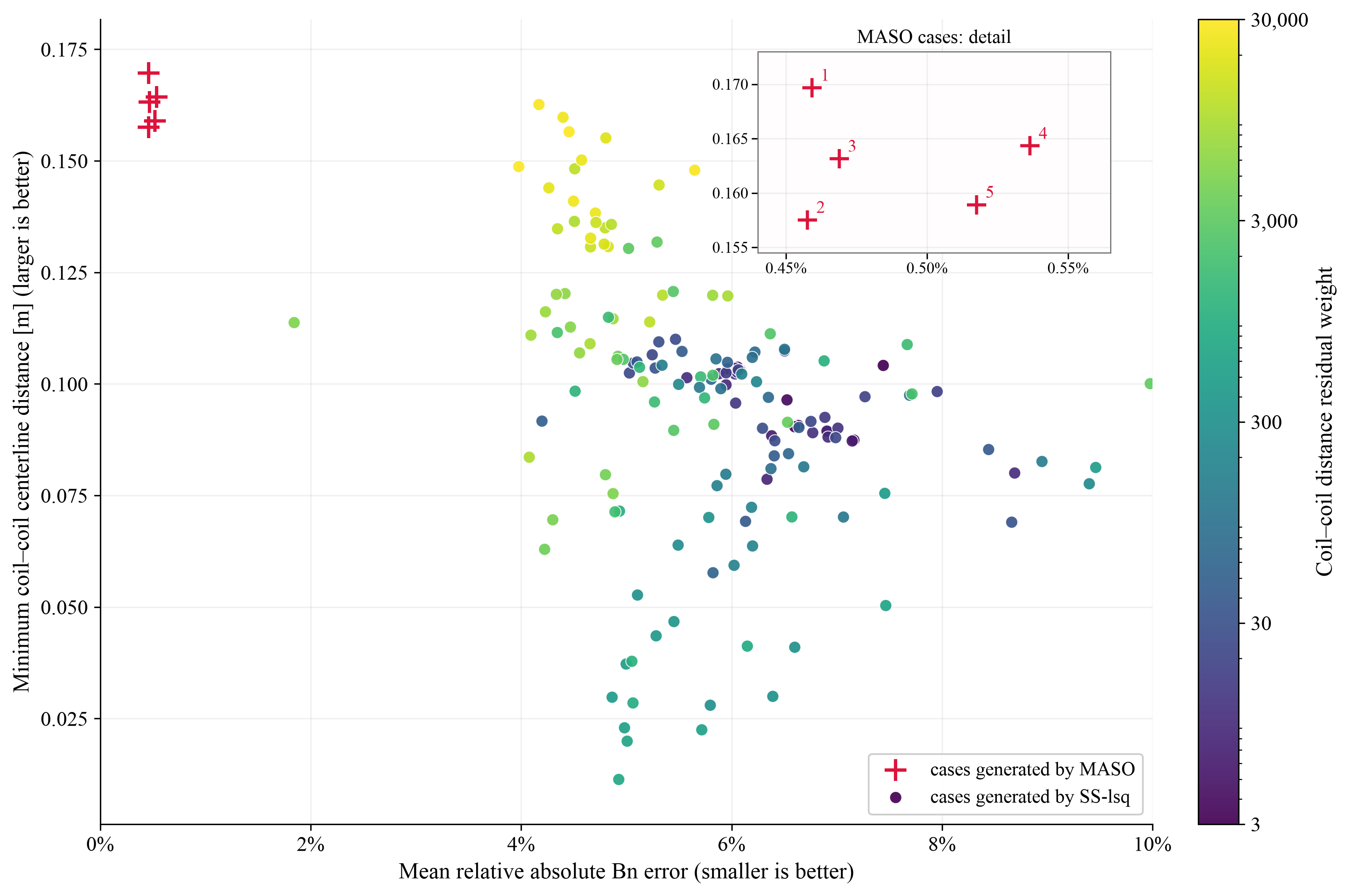}
    \caption{Pareto front of SS-lsq obtained by weight scanning, compared with the results of MASO.}
    \label{pareto}
\end{figure}

To demonstrate that MASO consistently outperforms SS-lsq, we performed an extensive scan over the weight combinations of SS-lsq and collected the resulting Pareto front in the trade-off between the normal-field error and minimum coil-coil distance, as shown in Figure~\ref{pareto}. Even at the best operating point of SS-lsq, both its residual normal-field error and its minimum coil-coil distance are larger than those obtained by MASO. This comparison strongly favors the proposed method.

\section{Summary and Discussion}

In this paper, a modified ADMM named MASO has been proposed for single-stage stellarator optimization of both plasma boundary and coils. To exploit the power of the algorithm, we have overcome the following difficulties: (i) By exploring the mathematical structure of the objective functions, we identify the variable-separable form of the objectives and exploit this structure to construct a new optimization paradigm. (ii) Since the constraints are nonlinear functions of both the surface and the coil variables, we extend the classical ADMM, which is originally formulated for linear constraints, to a nonlinear constraint version, in which the Lagrange multiplier and the penalty term are updated self-adaptively each round. (iii) Guided by the physical background of the objectives, we simplify the penalty terms, reducing them to the boundary normal field residual and a plasma-coil separation indicator function, which makes each subproblem feasible and practical.

The proposed method MASO has been applied to design modular coils for stellarators in this work. We started from an initial surface and coil set, and only provided the physics targets on the boundary and the engineering limits on the coils. The method successfully produced designs whose plasma performance and coil feasibility are better than those obtained by the traditional quasi-single stage or single-stage approaches. Furthermore, owing to the decoupling of the subproblems with lower dimensions, the method converges faster than the augmented Lagrangian method. Finally, when starting from a poor initial condition, MASO shows that the alternating scheme is of great robustness.

The constraint, the boundary normal field error, is the criterion for evaluating how well the coils reproduce the target boundary. Its explicit monitoring at every round provides direct information about the convergence of the constraints and could be used to control the required accuracy and to relax the tolerance during coil fabrication and assembly.

Moreover, MASO is not merely a single monolithic algorithm but a general plasma-coil co-optimization framework. In a plug-and-play manner, it can cooperate with existing stellarator optimization tools, so that the most suitable solver can be selected for each subproblem in a given scenario. This explicit decoupling endows MASO with a high degree of freedom. These properties, together with the efficiency demonstrated in the experiments of this work, make MASO a practical and extensible route to single-stage stellarator optimization.

In future work, we will further validate the effectiveness and generality of the proposed method by applying it to a wider variety of stellarator configurations and engineering requirements, such as quasi-axisymmetric and quasi helically symmetric designs with different numbers of field periods and configurations with more stringent constraints on the coil-coil and plasma-coil distance, curvature, and total coil length. Thanks to the plug-and-play nature of the framework, these extensions require only the substitution of the corresponding solvers and objectives, without reformulating the entire problem. Together with the efficiency and robustness demonstrated in this work, we therefore expect MASO to become a practical and general tool for the design of stellarators
with manufacturable coils.

\bibliographystyle{unsrt}
\bibliography{ref1, ref2}

\begin{thebibliography}{10}

\bibitem{liu_optimizing_2026}
Hengqian Liu, Guodong Yu, Caoxiang Zhu, José~Luis Velasco, Rahul Gaur, Dario
  Panici, Egemen Kolemen, Mingyang Yu, Weixing Ding, Shaojie Wang, and
  Ge~Zhuang.
\newblock Optimizing stellarators with hidden symmetry, March 2026.
\newblock arXiv:2502.09350 [physics.plasm-ph].

\bibitem{velasco_piecewise_2024}
J. L. Velasco, I.~Calvo, F. J. Escoto, E.~Sánchez, H.~Thienpondt, and
  F. I. Parra.
\newblock Piecewise {Omnigenous} {Stellarators}.
\newblock {\em Physical Review Letters}, 133(18):185101, October 2024.

\bibitem{landreman_omnigenity_2012}
Matt Landreman and Peter~J. Catto.
\newblock Omnigenity as generalized quasisymmetry.
\newblock {\em Physics of Plasmas}, 19(5):056103, May 2012.

\bibitem{liu_optimization_2026}
Hengqian Liu, Guodong Yu, José Luis~Velasco, and Caoxiang Zhu.
\newblock Optimization of stellarator configurations combining omnigenity and
  piecewise omnigenity.
\newblock {\em Nuclear Fusion}, 66(8):086043, August 2026.

\bibitem{helander_theory_2014}
Per Helander.
\newblock Theory of plasma confinement in non-axisymmetric magnetic fields.
\newblock {\em Reports on Progress in Physics}, 77(8):087001, August 2014.

\bibitem{dudt_magnetic_2024}
Daniel~W. Dudt, Alan~G. Goodman, Rory Conlin, Dario Panici, and Egemen Kolemen.
\newblock Magnetic {Fields} with {General} {Omnigenity}.
\newblock {\em Journal of Plasma Physics}, 90(1):905900120, February 2024.
\newblock arXiv:2305.08026 [physics.plasm-ph].

\bibitem{goodman_quasi-isodynamic_2024}
Alan~G. Goodman, Pavlos Xanthopoulos, Gabriel~G. Plunk, Håkan Smith, Carolin
  Nührenberg, Craig~D. Beidler, Sophia~A. Henneberg, Gareth Roberg-Clark,
  Michael Drevlak, and Per Helander.
\newblock Quasi-{Isodynamic} {Stellarators} with {Low} {Turbulence} as {Fusion}
  {Reactor} {Candidates}.
\newblock {\em PRX Energy}, 3(2):023010, June 2024.

\bibitem{helander_bootstrap_2009}
P~Helander and J~Nührenberg.
\newblock Bootstrap current and neoclassical transport in quasi-isodynamic
  stellarators.
\newblock {\em Plasma Physics and Controlled Fusion}, 51(5):055004, May 2009.

\bibitem{padidar_diffusion_2025}
Misha Padidar, Teresa Huang, Andrew Giuliani, and Marina Spivak.
\newblock Diffusion for {Fusion}: {Designing} {Stellarators} with {Generative}
  {AI}, November 2025.
\newblock arXiv:2511.20445 [cs].

\bibitem{gil_stellarator_2026}
Pedro~F. Gil, Weiping Li, Julianne Stratton, Alan~A. Kaptanoglu, and Eve~V.
  Stenson.
\newblock Stellarator {Coils} for {Future} {Fusion} {Reactors} via an
  {Augmented} {Lagrangian} {Approach}.
\newblock {\em Physical Review Letters}, 137(6):065101, August 2026.

\bibitem{kruger_coil_2025}
T.G. Kruger, M.F. Martin, D.A. Gates, and {the Thea Energy Team}.
\newblock Coil optimization methods for a planar coil stellarator.
\newblock {\em Nuclear Fusion}, 65(2):026051, February 2025.

\bibitem{yamaguchi_optimization_2021}
Hiroyuki Yamaguchi, Shinsuke Satake, Motoki Nakata, Akihiro Shimizu, Yasuhiro
  Suzuki, and The W7-X~Team.
\newblock Optimization of modular and helical coils applying genetic algorithm
  and fully-three-dimensional {B}-spline curves.
\newblock {\em Nuclear Fusion}, 61(10):106004, October 2021.

\bibitem{wu_planar_2025}
Ryan Wu, Thomas Kruger, and Charles Swanson.
\newblock Planar coil optimization for the {Eos} stellarator using sparse
  regression.
\newblock {\em Plasma Physics and Controlled Fusion}, 67(3):035019, March 2025.

\bibitem{nie_focus-hts_2025}
Xianyi Nie, Jianlin Peng, Yidong Xie, Guodong Yu, Ke~Liu, and Caoxiang Zhu.
\newblock {FOCUS}-{HTS}: a new stellarator coil design code for
  three-dimensional high-temperature superconducting magnets.
\newblock {\em Nuclear Fusion}, 65(8):086008, August 2025.

\bibitem{hammond_framework_2025}
K.~C. Hammond.
\newblock A framework for discrete optimization of stellarator coils.
\newblock {\em Nuclear Fusion}, 65(4):046012, April 2025.
\newblock arXiv:2412.00267 [physics].

\bibitem{zhu_designing_2018}
Caoxiang Zhu, Stuart~R Hudson, Yuntao Song, and Yuanxi Wan.
\newblock Designing stellarator coils by a modified {Newton} method using
  {FOCUS}.
\newblock {\em Plasma Physics and Controlled Fusion}, 60(6):065008, June 2018.

\bibitem{gil_efficient_2026}
Pedro~F. Gil, Weiping Li, Julianne Stratton, Alan~A. Kaptanoglu, and Eve~V.
  Stenson.
\newblock Efficient computation of stellarator coils with an augmented
  {Lagrangian} optimization method.
\newblock {\em Physical Review E}, 114(2):025202, August 2026.

\bibitem{henneberg_combined_2021}
S.~A. Henneberg, S.~R. Hudson, D.~Pfefferlé, and P.~Helander.
\newblock Combined plasma–coil optimization algorithms.
\newblock {\em Journal of Plasma Physics}, 87(2):905870226, April 2021.

\bibitem{giuliani_single-stage_2022}
Andrew Giuliani, Florian Wechsung, Antoine Cerfon, Georg Stadler, and Matt
  Landreman.
\newblock Single-stage gradient-based stellarator coil design: {Optimization}
  for near-axis quasi-symmetry.
\newblock {\em Journal of Computational Physics}, 459:111147, June 2022.

\bibitem{jorge_single-stage_2023}
R~Jorge, A~Goodman, M~Landreman, J~Rodrigues, and F~Wechsung.
\newblock Single-stage stellarator optimization: combining coils with fixed
  boundary equilibria.
\newblock {\em Plasma Physics and Controlled Fusion}, 65(7):074003, July 2023.

\bibitem{jorge_simplified_2024}
R.~Jorge, A.~Giuliani, and J.~Loizu.
\newblock Simplified and flexible coils for stellarators using single-stage
  optimization.
\newblock {\em Physics of Plasmas}, 31(11):112501, November 2024.

\bibitem{conlin_stellarator_2024}
Rory Conlin, Patrick Kim, Daniel~W. Dudt, Dario Panici, and Egemen Kolemen.
\newblock Stellarator optimization with constraints.
\newblock {\em Journal of Plasma Physics}, 90(5):905900501, October 2024.

\bibitem{bolte_proximal_nodate}
Jerome Bolte, Shoham Sabach, and Marc Teboulle.
\newblock Proximal {Alternating} {Linearized} {Minimization} for {Nonconvex}
  and {Nonsmooth} {Problems}.

\bibitem{boyd_distributed_2010}
Stephen Boyd.
\newblock Distributed {Optimization} and {Statistical} {Learning} via the
  {Alternating} {Direction} {Method} of {Multipliers}.
\newblock {\em Foundations and Trends® in Machine Learning}, 3(1):1--122,
  2010.

\bibitem{eckstein_douglasrachford_1992}
Jonathan Eckstein and Dimitri~P. Bertsekas.
\newblock On the {Douglas}—{Rachford} splitting method and the proximal point
  algorithm for maximal monotone operators.
\newblock {\em Mathematical Programming}, 55(1-3):293--318, April 1992.

\bibitem{bot_proximal_2020}
Radu~Ioan Bot and Dang-Khoa Nguyen.
\newblock The proximal alternating direction method of multipliers in the
  nonconvex setting: convergence analysis and rates, July 2020.
\newblock arXiv:1801.01994 [math].

\bibitem{li_global_2015}
Guoyin Li and Ting~Kei Pong.
\newblock Global convergence of splitting methods for nonconvex composite
  optimization.
\newblock {\em SIAM Journal on Optimization}, 25(4):2434--2460, January 2015.
\newblock arXiv:1407.0753 [math].

\bibitem{wang_global_2018}
Yu~Wang, Wotao Yin, and Jinshan Zeng.
\newblock Global {Convergence} of {ADMM} in {Nonconvex} {Nonsmooth}
  {Optimization}, May 2018.
\newblock arXiv:1511.06324 [math].

\bibitem{bian_stochastic_2021}
Fengmiao Bian, Jingwei Liang, and Xiaoqun Zhang.
\newblock A {Stochastic} {Alternating} {Direction} {Method} of {Multipliers}
  for {Non}-smooth and {Non}-convex {Optimization}.
\newblock {\em Inverse Problems}, 37(7):075009, July 2021.
\newblock arXiv:2012.07401 [math].

\bibitem{cui_convergence_2015}
Ying Cui, Xudong Li, Defeng Sun, and Kim-Chuan Toh.
\newblock On the convergence properties of a majorized {ADMM} for linearly
  constrained convex optimization problems with coupled objective functions,
  January 2015.
\newblock arXiv:1502.00098 [math.OC].

\bibitem{chan_plug-and-play_2016}
Stanley~H. Chan, Xiran Wang, and Omar~A. Elgendy.
\newblock Plug-and-{Play} {ADMM} for {Image} {Restoration}: {Fixed} {Point}
  {Convergence} and {Applications}, November 2016.
\newblock arXiv:1605.01710 [cs.CV].

\bibitem{ryu_plug-and-play_2019}
Ernest~K. Ryu, Jialin Liu, Sicheng Wang, Xiaohan Chen, Zhangyang Wang, and
  Wotao Yin.
\newblock Plug-and-{Play} {Methods} {Provably} {Converge} with {Properly}
  {Trained} {Denoisers}, May 2019.
\newblock arXiv:1905.05406 [cs.CV].

\bibitem{el_bourkhissi_convergence_2026}
Lahcen El~Bourkhissi and Ion Necoara.
\newblock Convergence rates for an inexact linearized {ADMM} for nonsmooth
  nonconvex optimization with nonlinear equality constraints.
\newblock {\em Computational Optimization and Applications}, 93(2):689--727,
  March 2026.

\bibitem{hien_multiblock_2023}
Le~Thi~Khanh Hien and Dimitri Papadimitriou.
\newblock Multiblock {ADMM} for nonsmooth nonconvex optimization with nonlinear
  coupling constraints, December 2023.
\newblock arXiv:2201.07657 [math.OC].

\bibitem{zhu_new_2018}
Caoxiang Zhu, Stuart~R. Hudson, Yuntao Song, and Yuanxi Wan.
\newblock New method to design stellarator coils without the winding surface.
\newblock {\em Nuclear Fusion}, 58(1):016008, January 2018.
\newblock arXiv:1705.02333 [physics].

\bibitem{glasser_generalized_2023}
A.~H. Glasser.
\newblock Generalized {Mercier} stability criterion for stellarators.
\newblock {\em Physics of Plasmas}, 30(5):052502, May 2023.

\bibitem{10.1063/1.864116}
S.~P. Hirshman and J.~C. Whitson.
\newblock Steepest‐descent moment method for three‐dimensional
  magnetohydrodynamic equilibria.
\newblock {\em The Physics of Fluids}, 26(12):3553--3568, 12 1983.

\bibitem{landreman_simsopt_2021}
Matt Landreman, Bharat Medasani, Florian Wechsung, Andrew Giuliani, Rogerio
  Jorge, and Caoxiang Zhu.
\newblock {SIMSOPT}: {A} flexible framework for stellarator optimization.
\newblock {\em Journal of Open Source Software}, 6(65):3525, September 2021.

\bibitem{panici_desc_2023}
Dario Panici, Rory Conlin, Daniel~W. Dudt, Kaya Unalmis, and Egemen Kolemen.
\newblock The {DESC} {Stellarator} {Code} {Suite} {Part} {I}: {Quick} and
  accurate equilibria computations, March 2023.
\newblock arXiv:2203.17173 [physics.plasm-ph].

\bibitem{conlin_desc_2023}
Rory Conlin, Daniel~W. Dudt, Dario Panici, and Egemen Kolemen.
\newblock The {DESC} {Stellarator} {Code} {Suite} {Part} {II}: {Perturbation}
  and continuation methods, April 2023.
\newblock arXiv:2203.15927 [physics.plasm-ph].

\bibitem{dudt_desc_2023}
Daniel Dudt, Rory Conlin, Dario Panici, and Egemen Kolemen.
\newblock The {DESC} {Stellarator} {Code} {Suite} {Part} {III}:
  {Quasi}-symmetry optimization.
\newblock {\em Journal of Plasma Physics}, 89(2):955890201, April 2023.
\newblock arXiv:2204.00078 [physics.plasm-ph].

\end{thebibliography}

\newpage
\appendix

\end{document}